\documentclass{article}
\usepackage{graphicx} % Required for inserting images
\usepackage{paralist}
\usepackage{enumerate} 
\usepackage{dcolumn}
\usepackage[OT1]{fontenc}
\usepackage[numbers]{natbib}
\usepackage{booktabs}
\usepackage[colorlinks,citecolor=blue,urlcolor=blue]{hyperref}
\usepackage{amsmath,amssymb,amsthm,amscd,graphicx,color,accents}
\usepackage[margin=2cm]{geometry}

\usepackage{caption}
\usepackage{placeins}

\def\bX{{\boldsymbol{X}}}
\def\bY{{\boldsymbol{Y}}}

\def\bzero{\boldsymbol{0}}

\def\P{{\mathbb P}}

\def\U{{\mathcal U}}
\def\V{{\mathcal V}}

\def\N{\mathbb{N}}
\def\R{\mathbb{R}}

\def\mX{\mathcal{X}}
\def\mY{\mathcal{Y}}

\def\mB{\mathcal{B}}
\def\mZ{\mathcal{Z}}

\def\blam{\boldsymbol{\lambda}}
\def\bmu{\boldsymbol{\mu}}

\newtheorem{theorem}{Theorem}[section]
\newtheorem{proposition}[theorem]{Proposition}

\newtheorem{corollary}[theorem]{Corollary}
\newtheorem{remark}[theorem]{Remark}

\title{\bf Approximating the Null Distribution of\\ Generalized Distance Covariance}

\author{Dominic Edelmann\thanks{Division of Biostatistics, German Cancer Research Center,
Im Neuenheimer Feld 280, 69120 Heidelberg, Germany.
E-mail address: \href{mailto:dominic.edelmann@dkfz-heidelberg.de}{dominic.edelmann@dkfz-heidelberg.de}.}}

\date{}

\begin{document}

\maketitle

\section{Introduction}

First proposed by Sz\'{e}kely and his coauthors \cite{szekely2007measuring}, distance covariance is now widely regarded as a standard tool for assessing independence of random variables. In practice, distance covariance testing is mostly conducted using permutation procedures. However, permutation methods can become computationally demanding for large sample sizes or when extremely small p-values must be estimated with high precision \cite{edelmann2025generalized}. Two- \cite{HH} and three-moment \cite{berschneider2018complex} matching approaches as well as conservative approximations based on simple parametric families \cite{shen2022chi,berschneider2018complex} have been proposed as computationally efficient alternatives; however these approximations can exhibit substantial discrepancies in the tails of the distribution.

A natural approach is to approximate the asymptotic null distribution directly, which is known to be a weighted sum of chi-square distributed random variables \cite{szekely2007measuring}. Although this strategy is less commonly used for distance covariance, it is more prevalent in the context of the Hilbert--Schmidt Independence Criterion (HSIC), which can be viewed as a kernel-based generalization of distance covariance \cite{sejdinovic2013equivalence,edelmann2022regression}. While several implementations of this idea have been proposed and partial results are available \cite{gretton2009fast}, a comprehensive theoretical justification of this approach is still lacking.

In this work, we close this gap by rigorously establishing the validity of this direct approximation under mild and transparent assumptions. This result is derived for a generalization of distance covariance presented in this article, which is closely related to the general distance covariance presented in \cite{sejdinovic2013equivalence,edelmann2022regression}. In order to obtain a fully rigorous treatment, particular attention is paid to measurability and separability issues, which are often left implicit in the existing literature.

In addition to this theoretical result, we develop approaches to reduce computation time and improve precision of the p-value approximation. In particular, we show that a restricted set of eigenvalues allows for establishing upper and lower bounds for the p-value, which usually rapidly converge in practice. This reduces the complexity from $O(n^3)$ operations that are needed to evaluate the full spectra of two $n \times n$ matrices to $O(k n^2)$ operations that are required to evaluate the $k$ largest eigenvalues using a Lanczos algorithm. Moreover, we propose a shrinkage approach for the weights of the corresponding Gaussian quadratic form, matching the second moment of the form to an unbiased estimate of the second moment of the test statistic, which is available in closed form at cost $O(n^2)$ \cite{berschneider2018complex}.

Simulations under the null hypothesis of independence demonstrate that the derived tests are - apart from costly Monte Carlo approaches - the only available tests for which the empirical type I error converges to the nominal level. In general, the novel tests dominate existing approaches for moderate sample sizes ($n \geq 100$), while they do show some problems with very small sample sizes. The simulations also show that the proposed shrinkage approach clearly outperforms the naive method without shrinkage.

The outline of this article is as follows. In Section~\ref{sec:clt}, we introduce our version of generalized distance covariance and derive its asymptotic distribution under independence. In Section~\ref{sec:main}, we derive the validity of the spectral approach for independence testing with generalized distance covariance. Section~\ref{sec:implementation} describes the computational implementation, and Section~\ref{sec:simulations} provides a set of simulations comparing the performance of the spectral approach and its competitors. The accompanying Supplementary Material contains proofs for all theoretical results and additional details on the proposed algorithms.

\section{The asymptotic distribution of generalized distance covariance in the case of independence} \label{sec:clt}

Let $\mathcal{Z}$ denote a set. Throughout this work, a \textit{distance} will be defined as a function $d : \mZ  \times \mZ \to [0,\infty)$ satisfying, for $x,y \in \mZ$:
\begin{enumerate}
    \item[(S)] \textit{Symmetry:} $d(x,y) = d(y,x)$,
    \item[(Z)] \textit{Zero self-distance:} $d(x,x) = 0$.
\end{enumerate}

A \textit{distance} is called a \textit{metric} if it additionally satisfies the following two conditions for $x,y,z \in \mZ$
\begin{enumerate}
   \item[(I)] \textit{Identity of indiscernibles:} $d(x,y) = 0$ $\Rightarrow$ $x = y$,
    \item[(T)] \textit{Triangle inequality:} $d(x,z) \leq d(x,y) + d(y,z)$.
\end{enumerate}
%A distance  satisfying condition (I) will be called a I-distance, a distance satisfying condition (T) will be called a T-distance.

We further say that a distance $d : \mZ  \times \mZ \to [0,\infty)$ is of negative type if for all $n \geq 2$, $z_1,\ldots,z_n \in \mZ$ and $a_1,\ldots,a_n \in \R$ with $\sum_{i=1}^n a_i = 0$,
		$$
			\sum_{i,j=1}^n a_i a_j d(z_i,z_j) \leq 0 .
		$$

Let $(\mathcal{X},\rho_\mX)$ and $(\mY,\rho_\mY)$ denote two separable metric spaces equipped with the corresponding Borel $\sigma$-algebras $\mB_\mX$ and $\mB_\mY$.  Moreover, let $d_\mathcal{X} :\mathcal{X} \times \mathcal{X} \to [0,\infty)$ and   $d_\mathcal{Y} :\mathcal{Y} \times \mathcal{Y} \to [0,\infty)$ denote distances of negative type that are continuous with respect to the corresponding product topologies. Note that this includes the important special case where $d_\mX = \rho_\mX$ and $d_\mY  = \rho_\mY$. The measurability and continuity conditions given in this paragraph are typically omitted in the literature (see e.g. \cite{sejdinovic2013equivalence, edelmann2022regression}); however it appears that these conditions are required to develop many important results, such as e.g. the central limit theorem given in Theorem \ref{th:asydist}.

In the following, we will assume that $X$ and $Y$ denote jointly distributed random variables  with values in  $\mathcal{X}$ and $\mathcal{Y}$  and that their distributions are given by $P^X$ and $P^Y$, respectively. For all $(X,Y)$ such that
$$
\int d_\mathcal{X}(x_1,x_2)  dP^{X}(x_1) dP^{X}(x_2) < \infty, \quad \int d_\mathcal{Y}(y_1,y_2)  dP^{Y}(y_1) dP^{Y}(y_2)  < \infty,
$$
the generalized distance covariance is defined as:
\begin{align}
		&\V^2(d_\mX,d_\mY;\, X,Y) \nonumber \\&=  \int \left\{ d_\mX(x_1,x_2) -d_\mX(x_3,x_2) -d_\mX(x_1,x_4)+d_\mX(x_3,x_4)\right\} \nonumber \\
     &\quad  \quad \times \left\{d_\mY(y_1,y_2)-d_\mY(y_5,y_2) -d_\mY(y_1,y_6)+d_\mY(y_5,y_6)\right\} \bigotimes_{i=1}^6 dP^{(X,Y)}  (x_i,y_i),
     \label{eq:gendcov2}
\end{align}
where $dP^{(X,Y)}$ denotes the joint probability measure of $(X,Y)$ and $\bigotimes_{i=1}^m P_i$ is the product measure of $(P_1,\ldots,P_m)$.
Now define the unsymmetrized kernel $g$ as

 \begin{align}
		& g((X_{i_1},Y_{i_1}),\ldots,(X_{i_6},Y_{i_6})) =  \nonumber \\&=  \left\{ d_\mX(X_{i_1},X_{i_2}) -d_\mX(X_{i_3},X_{i_2}) -d_\mX(X_{i_1},X_{i_4})+d_\mX(X_{i_3},X_{i_4})\right\} \nonumber \\
     &\quad  \quad \times \left\{d_\mY(Y_{i_1},Y_{i_2})-d_\mY(Y_{i_5},Y_{i_2}) -d_\mY(Y_{i_1},Y_{i_6})+d_\mY(Y_{i_5},Y_{i_6})\right\}.
     \label{eq:kernel}
\end{align}
Denoting by $\Pi_6$ the set of all permutations of $\{1,\ldots,6\}$, its symmetrized version is given by
\begin{equation} \label{eq:h}
   h((X_{i_1},Y_{i_1}),\ldots,(X_{i_6},Y_{i_6})) = \frac{1}{6!} \sum_{\pi \in \Pi_6} g \big( (X_{i_{\pi(1)}},Y_{i_{\pi(1)}}),\ldots,(X_{i_{\pi(6)}},Y_{i_{\pi(6)}}) \big).
\end{equation}
The natural approach for estimating $\V^2(d_\mX,d_\mY;\, X,Y) $ is employing the U-statistic \cite{HH}
   $$
        \widehat{\U}_n^2(\bX,\bY) = {n \choose 6}^{-1} \sum_{1 \leq i_1 < i_2 < \cdots < i_6 \leq n} h((X_{i_1},Y_{i_1}),\ldots,(X_{i_6},Y_{i_6})),
    $$
or the corresponding V-statistic \cite{sejdinovic2013equivalence, bottcher2018detecting}
   $$
        \widehat{\V}_n^2(\bX,\bY) = \frac{1}{n^6} \sum_{i_1,\ldots, i_6=1}^n h((X_{i_1},Y_{i_1}),\ldots,(X_{i_6},Y_{i_6})).
    $$

With analogous arguments as given for the ordinary distance covariance in \cite{fokianos2018testing}, one can show that
$$
  \widehat{\V}_n^2(\bX,\bY) = \frac{1}{n^2} \operatorname{tr} (A_n B_n)
  = \frac{1}{n^2} \sum_{i,j=1}^n (A_n)_{ij} (B_n)_{ij},
$$
where $A_n$ and $B_n$ are the $n \times n$ matrices with entries

\begin{align}
 (A_{n})_{ij } &= d_\mX (X_i, X_j) - \frac{1}{n} \sum_{k=1}^n d_\mX (X_i, X_k) - \frac{1}{n} \sum_{l=1}^n d_\mX (X_l, X_j) +  \frac{1}{n^2} \sum_{l,k=1}^n d_\mX (X_l, X_k), \nonumber \\
  (B_{n})_{ij } &= d_\mY (Y_i, Y_j) - \frac{1}{n} \sum_{k=1}^n d_\mY (Y_i, Y_k) - \frac{1}{n} \sum_{l=1}^n d_\mY (Y_l, Y_j) +  \frac{1}{n^2} \sum_{l,k=1}^n d_\mY (Y_l, Y_k),
  \label{eq:AB}
\end{align}
a similar $O(n^2)$ representation holds for $\widehat{\U}_n^2(\bX,\bY)$ \cite{huo2016fast}.

After these preliminaries, we can now state the central limit theorem for distance covariance as required in the remainder of this paper. Similar results appear elsewhere (see e.g., \cite{jakobsen2017distance}), though our version assumes the general class of distances defined above together with the stated measurability and continuity conditions, and so does not coincide exactly with those statements.

\begin{theorem} \label{th:asydist}
Let $X$ and $Y$ be independent and 
\begin{align*}
&0 < \int d_\mathcal{X}(x_1,x_2)  dP^X(x_1) dP^{X}(x_2) < \infty, \\ \quad &0 < \int d_\mathcal{Y}(y_1,y_2)  dP^{Y}(y_1) dP^{Y}(y_2)  < \infty.
\end{align*}
Then, for $n \to \infty$, 
    $$
   n \, \widehat{\U}_n^2(\bX,\bY) \stackrel{\mathcal{D}}{\longrightarrow} \sum_{i,j=1}^\infty \lambda^X_i \lambda^Y_j (Z_{ij}^2-1),
    $$
and, 
    $$
   n \, \widehat{\V}_n^2(\bX,\bY) \stackrel{\mathcal{D}}{\longrightarrow} \sum_{i,j=1}^\infty \lambda^X_i \lambda^Y_j Z_{ij}^2,
    $$
where $(Z_{ij})_{i,j \in \N}$ is a family of i.i.d.\ standard normal random variables, $\lambda^X_1 \geq \lambda^X_2 \geq \cdots \geq 0$ with $\lambda^X_1 > 0$ are the eigenvalues of the operator $T^{X}: L^2(\mX, P^X) \to L^2(\mX,P^X)$, 
    $$
(T^{X} f  ) (x_1) = \int D^X (x_1,x_2) f(x_2) dP^X (x_2),
    $$
and  $\lambda^Y_1 \geq \lambda^Y_2 \geq \cdots \geq 0$ with $\lambda^Y_1 > 0$ are the eigenvalues of the operator $T^{Y}: L^2(\mY, P^Y) \to L^2(\mY, P^Y)$,
    $$
(T^{Y} f  ) (y_1)  = \int D^Y (y_1,y_2) f(y_2) dP^Y (y_2),
    $$
where
$$
D^X(x_1,x_2) = \int (-d_\mX (x_1,x_2) + d_\mX  (x_3,x_2)  + d_\mX  (x_1,x_4) -  d_\mX  (x_3,x_4)) d P^X (x_3) d P^X (x_4)
$$
and 
$$
D^Y(y_1,y_2) =  \int (-d_\mY  (y_1,y_2) + d_\mY (y_3,y_2)  + d_\mY (y_1,y_4) -  d_\mY(y_3,y_4)) d P^Y (y_3) d P^Y (y_4).
$$
\end{theorem}

\section{Approximating the asymptotic distribution} \label{sec:main}

Throughout this section, weight sequences are taken from
\[
	\ell^1_{\downarrow} := \Big\{ \bmu = (\mu_1,\mu_2,\ldots) \, : \,
	\mu_1 \geq \mu_2 \geq \cdots \geq 0, \ \textstyle\sum_{i=1}^\infty \mu_i < \infty \Big\},
\]
a finite vector $(\mu_1,\ldots,\mu_m)$ being identified with the sequence obtained by setting $\mu_i = 0$ for $i > m$. For $\bmu^X, \bmu^Y \in \ell^1_{\downarrow}$ and $(Z_{ij})_{i,j \in \N}$ a family of i.i.d.\ standard normal random variables, we define, for $c \in \R$,
\begin{align}
	F(c; \bmu^X, \bmu^Y) &:= \P \Big( \sum_{i,j=1}^{\infty} \mu^X_i \mu^Y_j (Z_{ij}^2-1) \leq c \Big), \label{eq:F} \\
	G(c; \bmu^X, \bmu^Y) &:= \P \Big( \sum_{i,j=1}^{\infty} \mu^X_i \mu^Y_j Z_{ij}^2 \leq c \Big). \label{eq:G}
\end{align}
Both series converge almost surely and in $L^1$, since $\sum_{i,j=1}^\infty \mu^X_i \mu^Y_j = \big(\sum_{i=1}^\infty \mu^X_i\big)\big(\sum_{j=1}^\infty \mu^Y_j\big) < \infty$; hence $F$ and $G$ are well defined. For $u \in (0,1)$ we write
$$
	F^{-1}(u; \bmu^X, \bmu^Y) := \inf \{ c \in \R \, : \, F(c; \bmu^X, \bmu^Y) \geq u \},
$$
and analogously for $G$.

Let now $(X_1,Y_1), \ldots, (X_n,Y_n)$ be i.i.d.\ copies of $(X,Y)$ and let $A_n$ and $B_n$ be the doubly centred distance matrices defined in \eqref{eq:AB}. Since $d_\mX$ and $d_\mY$ are of negative type, the matrices $-n^{-1} A_n$ and $-n^{-1}B_n$ are positive semi-definite; moreover $A_n \boldsymbol{1} = B_n \boldsymbol{1} = \bzero$, so that both have rank at most $n-1$. We denote their eigenvalues by
$$
	\widehat{\lambda}^X_{n1} \geq \cdots \geq \widehat{\lambda}^X_{n,n-1} \geq \widehat{\lambda}^X_{nn} = 0
	\qquad \text{and} \qquad
	\widehat{\lambda}^Y_{n1} \geq \cdots \geq \widehat{\lambda}^Y_{n,n-1} \geq \widehat{\lambda}^Y_{nn} = 0,
$$
respectively, and set $\widehat{\lambda}^X_{nm} = \widehat{\lambda}^Y_{nm} = 0$ for $m > n$, so that
$\widehat{\blam}^X_n := (\widehat{\lambda}^X_{nm})_{m \in \N}$ and $\widehat{\blam}^Y_n := (\widehat{\lambda}^Y_{nm})_{m \in \N}$
are elements of $\ell^1_{\downarrow}$. Writing $\blam^X := (\lambda^X_i)_{i \in \N}$ and $\blam^Y := (\lambda^Y_j)_{j \in \N}$ for the population eigenvalue sequences of Theorem~\ref{th:asydist}, which lie in $\ell^1_{\downarrow}$ by the trace identities established in its proof, we set
\begin{alignat*}{3}
	\widehat{F}_n(\cdot) &:= F(\cdot \, ; \widehat{\blam}^X_n, \widehat{\blam}^Y_n), &\qquad
	\widehat{G}_n(\cdot) &:= G(\cdot \, ; \widehat{\blam}^X_n, \widehat{\blam}^Y_n), \\
	F_0(\cdot) &:= F(\cdot \, ; \blam^X, \blam^Y), &\qquad
	G_0(\cdot) &:= G(\cdot \, ; \blam^X, \blam^Y).
\end{alignat*}
Note that $\widehat{F}_n$ and $\widehat{G}_n$ are random distribution functions, depending on the sample only through the spectra of $A_n$ and $B_n$.

\begin{theorem} \label{th:approx}
	Assume that
\begin{align}
&0 < \int d_\mathcal{X}(x_1,x_2)  dP^X(x_1) dP^{X}(x_2) < \infty, \nonumber\\   &0 < \int d_\mathcal{Y}(y_1,y_2)  dP^{Y}(y_1) dP^{Y}(y_2)  < \infty . \label{eq:moments}
\end{align}
	Then there exists an  $\Omega_0$ with $\P(\Omega_0) = 1$ such that, on $\Omega_0$, 
	$$
		\sup_{c \in \R} \big| \widehat{F}_n(c) - F_0(c) \big| \longrightarrow 0
		\qquad \text{and} \qquad
		\sup_{c \in \R} \big| \widehat{G}_n(c) - G_0(c) \big| \longrightarrow 0 .
	$$
\end{theorem}

Note that Theorem~\ref{th:approx} does not require $X$ and $Y$ to be independent: the spectra of $A_n$ and $B_n$ are consistent for the marginal population spectra irrespective of the dependence structure between $X$ and $Y$. Combining Theorem~\ref{th:approx} with Theorem~\ref{th:asydist} yields the asymptotic validity of the resulting tests.

\begin{corollary} \label{cor:level}
	Let \eqref{eq:moments} hold, and let $X$ and $Y$ be independent. Then, for every $\alpha \in (0,1)$, as $n \to \infty$,
	$$
		\P \big( n \, \widehat{\U}_n^2(\bX,\bY) > \widehat{F}^{-1}_n(1-\alpha) \big) \longrightarrow \alpha
		\qquad \text{and} \qquad
		\P \big( n \, \widehat{\V}_n^2(\bX,\bY) > \widehat{G}^{-1}_n(1-\alpha) \big) \longrightarrow \alpha .
	$$
\end{corollary}

\section{Fast and precise computational implementation} \label{sec:implementation}

Evaluating $\widehat{G}_n$ (or $\widehat{F}_n$, respectively) requires the eigenvalues of two $n \times n$ matrices, which usually come at a cost of $O(n^3)$. This section describes an implementation that avoids the full decomposition whenever the desired accuracy permits. In addition, we propose a shrinkage approach for the Gaussian quadratic form corresponding to $\widehat{G}_n$, matching the first two moments of $\widehat{G}_n$ to unbiased estimates of the moments of the test statistic $n \widehat\V_n^2$.

The distribution $\widehat{G}_n$ is that of
\begin{equation} \label{eq:finite_spectral}
    T_n = \sum_{i,j=1}^{n-1} \widehat\lambda^X_{ni} \widehat\lambda^Y_{nj} \, Z_{ij}^2 .
\end{equation}

We will denote the observed statistic by $t_{\mathrm{obs}} = n \widehat{\V}_n^2(\bX,\bY)$. Moreover, for the remainder of this section, we will use the notation
\begin{equation} \label{eq:mom1}
    m_1 = \sum_{i,j=1}^{\infty} \widehat\lambda^X_{ni} \widehat\lambda^Y_{nj} = \frac{a_{\cdot\cdot} \, b_{\cdot\cdot}}{n^4}, \qquad a_{\cdot\cdot} = \sum_{i,j=1}^n d_\mX(X_i,X_j), \quad b_{\cdot\cdot} = \sum_{i,j=1}^n d_\mY(Y_i,Y_j) .
\end{equation}

\subsection{Adaptive eigenvalue computation using bounds on the $p$-value} \label{sec:bounds}

Suppose the $k$ largest eigenvalues of each centered matrix have been computed (e.g., by a Lanczos method), yielding the $k^2$ products $\widehat\lambda^X_{ni} \widehat\lambda^Y_{nj}$, $1 \leq i,j \leq k$, which we collect in non-increasing fashion in the vector $\boldsymbol\lambda^{k}$. By \eqref{eq:mom1}, the sum of the omitted eigenvalues
\begin{equation} \label{eq:rest}
    R = m_1 - \Big( \sum_{i=1}^k \widehat\lambda^X_{ni} \Big) \Big( \sum_{j=1}^k \widehat\lambda^Y_{nj} \Big) \geq 0
\end{equation}
is known without any further eigenvalue computation. We write
$$
    p(\boldsymbol\lambda) := \P \Big( \sum_i \lambda_i Z_i^2 > t_{\mathrm{obs}} \Big)
$$
for the upper tail probability associated with a vector $\boldsymbol\lambda$ of non-negative weights, where $(Z_i)_{i \in \N}$ denotes a family of i.i.d.\ standard normal random variables, so that the p-value of interest is $p(\widehat{\boldsymbol\lambda})$ for the full vector $\widehat{\boldsymbol\lambda}$  collecting $( \widehat\lambda^X_{ni} \widehat\lambda^Y_{nj} )_{1 \leq i,j \leq n-1}$ in non-increasing fashion.

Since $\widehat{\boldsymbol\lambda}$ arises from $\boldsymbol\lambda^{k}$ by adjoining non-negative weights, we obtain the elementary lower bound
\begin{equation} \label{eq:lower}
    p_{\mathrm{anti}} := p(\boldsymbol\lambda^{k}) \, \leq \, p(\widehat{\boldsymbol\lambda}).
\end{equation}

For obtaining an upper bound, we require the notion of weak majorization \cite{marshall1979,tong1980}. One says that $\boldsymbol\lambda$ \emph{weakly majorises} $\boldsymbol\mu$ if their non-increasing rearrangements satisfy $\sum_{i \leq m} \lambda_i \geq \sum_{i \leq m} \mu_i$ for every $m$. The following proposition is a direct consequence of \cite[Corollary 3]{szekely2003extremal}.

\begin{proposition} \label{prop:majorisation}
    Let $\boldsymbol\lambda$ and $\boldsymbol\mu$ be vectors of non-negative weights such that $\boldsymbol\lambda$ weakly majorises $\boldsymbol\mu$, and let $\kappa = \sum_i \mu_i$. Then, for every $t \geq 2 \kappa$,
    $$
        \P \Big( \sum_i \lambda_i Z_i^2 > t \Big) \geq \P \Big( \sum_i \mu_i Z_i^2 > t \Big) .
    $$
\end{proposition}

When only observing $\boldsymbol\lambda^{k}$, we know that every omitted eigenvalue is bounded by
\begin{equation} \label{eq:val}
    v := \max \big( \widehat\lambda^X_{n1} \widehat\lambda^Y_{nk}, \, \widehat\lambda^Y_{n1} \widehat\lambda^X_{nk} \big) \, \geq \, \widehat\lambda^X_{ni} \widehat\lambda^Y_{nj}, \qquad \max(i,j) > k .
\end{equation}
Defining $\boldsymbol\lambda^{\mathrm{cons}}$ as
$$
    \boldsymbol\lambda^{\mathrm{cons}} := ( \boldsymbol\lambda^{k}, \, v^{(d)}, \, r ), \qquad d = \lfloor R/v \rfloor, \quad r = R - dv,
$$
where $v^{(d)}$ denotes the value $v$ repeated $d$ times,
it hence follows that $\boldsymbol\lambda^{\mathrm{cons}}$ weakly majorises $\widehat{\boldsymbol\lambda}$. Moreover, one easily checks that the sum of the elements of $\boldsymbol\lambda^{\mathrm{cons}}$ is $m_1$; we hence obtain

\begin{corollary} \label{cor:bounds}
    For $t_{\mathrm{obs}} \geq 2 m_1$,
    $$
        p(\widehat{\boldsymbol\lambda}) \, \leq \, p_{\mathrm{cons}} := p(\boldsymbol\lambda^{\mathrm{cons}}) .
    $$
\end{corollary}

Hence, for $t_{\mathrm{obs}} \geq 2 m_1$, calculating the first $k$ eigenvalues of $-n^{-1} A_n$ and $-n^{-1} B_n$ provides upper and lower bounds for the p-value of interest,
$$
p_{\mathrm{anti}} \leq p(\widehat{\boldsymbol\lambda}) \leq \, p_{\mathrm{cons}}.
$$
Whenever the desired precision for the p-value is achieved (e.g. $p_{\mathrm{cons}}/p_{\mathrm{anti}} \leq \tau_\mathrm{tol}$ for some prespecified $\tau_\mathrm{tol} > 1$) or we are certain that our test does or does not reject the null at a certain significance level, we may stop calculating further eigenvalues and output $p_{\mathrm{cons}}$ (or another reasonable choice in the interval $[p_{\mathrm{anti}}, p_{\mathrm{cons}}])$. If this is not the case, we can continue and calculate a larger set of eigenvalues in the next step.

The upper bound $p_{\mathrm{cons}}$ is not applicable for $t_{\mathrm{obs}} < 2 m_1$. We propose to handle this issue via two mechanisms. First, since $t_{\mathrm{obs}} < 2 m_1$ mostly corresponds to a large p-value and the precise evaluation of large p-values is rarely of interest, we propose to stop the evaluation of further eigenvalues, when we can be certain that the p-value is greater than a certain prespecified value $\tau_{\mathrm{large}}$, i.e., whenever $p_{\mathrm{anti}} > \tau_{\mathrm{large}}$. Second, we propose to stop the evaluation, when the sum of the remaining non-identified eigenvalues is small, say $R/m_1 < \tau_\mathrm{conv}$. To correct the first moment of the test-statistic, we then propose to calculate the p-value based on the distribution of
\begin{equation} \label{eq:stattrunc}
\sum_{i,j=1}^{k} \widehat\lambda^X_{ni} \widehat\lambda^Y_{nj} \, Z_{ij}^2 + R.
\end{equation}

In conclusion, starting from $k = k_0$, the proposed algorithm is

\begin{enumerate}
    \item Calculate the first $k$ eigenvalues of $-n^{-1} A_n$ and $-n^{-1} B_n$ using a Lanczos method and evaluate the bounds described above.
    \item If $p_{\mathrm{anti}} > \tau_{\mathrm{large}}$, the result is clearly non-significant and $p_{\mathrm{anti}}$ is returned;
\item \begin{enumerate} \item Case 1: $t_{\mathrm{obs}} \geq 2 m_1$.
 \quad If $p_{\mathrm{cons}} / p_{\mathrm{anti}} \leq \tau_{\mathrm{tol}}$ and both bounds lie on the same side of a nominal level $\alpha$ supplied by the user, the bracket is sufficiently tight and $p_{\mathrm{cons}}$ is returned;
    \item Case 2: $t_{\mathrm{obs}} < 2 m_1$.
 \quad If $R/m_1 < \tau_\mathrm{conv}$, the distribution of the test statistic is sufficiently well approximated; the p-value is calculated using the statistic in \eqref{eq:stattrunc}.
    \end{enumerate} 
    \item otherwise $k$ is replaced by $\min(\tau_{\mathrm{mult}} k, n)$ and the procedure is repeated from step 1.
\end{enumerate}

Once $k$ exceeds $\lfloor 0.15 n \rfloor$, the scheme is abandoned. Beyond this point a full eigendecomposition is - from our experience - no more expensive than the next iteration.

\subsection{Exact moments and eigenvalue shrinkage} \label{sec:shrinkage}

A disadvantage of the test procedure proposed above is that the moments of the distribution $\widehat{G}_n$ are biased estimators of the moments of the test statistic $n \widehat{\V}_n^2$.

To obtain a more precise evaluation of the finite sample distribution, we propose to modify the distribution $\widehat{G}_n$ to match unbiased estimates of the first and second moments. In \cite[Theorem 4.15]{berschneider2018complex}, unbiased estimators for the first two finite sample moments have been derived. In particular, an unbiased estimator for the first moment of $n \widehat{\V}_n^2$ is given by
    $$
     \widetilde{m}_1 = a_{\cdot\cdot} b_{\cdot\cdot} / \{ n^3 (n-1) \}.
    $$
 The unbiased estimator for the second moment $\widetilde{m}_2$ of $n \widehat{\V}_n^2$ admits a rather lengthy - but $O(n^2)$-computable - expression in terms of the quantities $\sum_{i,j} d_\mX(X_i,X_j)^2$, $\sum_i \big( \sum_j d_\mX(X_i,X_j) \big)^2$ and $a_{\cdot\cdot}$, and their counterparts for $\mY$, see \cite[Eq. (4.42)]{berschneider2018complex} or the corresponding implementation in \texttt{dcortools} for details.

%When the full decomposition is computed, the product weights are exact functions of the sample distance matrices but remain noisy estimates of the population products $\lambda^X_i \lambda^Y_j$, sample eigenvalues being over-dispersed relative to their population counterparts. This can be corrected using exact moments of $T_n$.

Modifying $\widehat{G}_n$ to match the first moment $\widetilde{m}_1$ can be conveniently done by using the eigenvalues of 
$$
    - \{ n(n-1) \}^{-1/2} A_n, \qquad - \{ n(n-1) \}^{-1/2} B_n ,
$$
instead of those of $-n^{-1} A_n$ and $-n^{-1} B_n$. 

Matching to the unbiased estimate of the second moment is more involved and only performed in cases where the full eigendecomposition has been evaluated. With a slight abuse of notation, let $\widehat\lambda^X_{ni}$ and $\widehat\lambda^Y_{nj}$, $i,j = 1,\ldots,n-1$ denote these eigenvalues and let $\widehat\lambda_{ij} = \widehat\lambda^X_{ni} \widehat\lambda^Y_{nj}$. Then, noting that the variance of a chi-square distribution with $k$ degrees of freedom is $2k$, the variance of 
$$
\sum_{i,j=1}^{n-1} \widehat\lambda_{ij} \, Z_{ij}^2
$$
is $2 \, \sum_{i,j} \widehat\lambda_{ij}^{\,2}$.
Write  $\bar\lambda = (n-1)^{-2} \widetilde{m}_1$ for their mean, and
$$
    \widetilde{\sigma}^2 = \tfrac12 \big( \widetilde{m}_2 - \widetilde{m}_1^2 \big) 
$$
for the target sum of squares obtained from the unbiased moment estimates. In most cases, the sample eigenvalues will be overdispersed compared to their
population counterparts, and hence $\sum_{i,j} \widehat{\lambda}_{ij}^{\,2} > \widetilde{\sigma}^2 > (n-1)^2 \, \bar\lambda^2$.
For these cases, we propose to replace the eigenvalues $\widehat\lambda_{ij}$ by
their shrinkage counterparts
\begin{equation} \label{eq:shrinkage}
    \widetilde\lambda_{ij} = \alpha \, \widehat\lambda_{ij} + (1-\alpha) \, \bar\lambda, \qquad
    \alpha = \bigg( \frac{\widetilde{\sigma}^2 - (n-1)^2 \, \bar\lambda^2}{\sum_{i,j} (\widehat\lambda_{ij} - \bar\lambda)^2} \bigg)^{1/2}.
\end{equation}
It then holds:
\begin{proposition} \label{prop:shrinkage}
    Let $\sum_{i,j} \widehat{\lambda}_{ij}^{\,2} > \widetilde{\sigma}^2 > (n-1)^2 \, \bar\lambda^2$. Then
    $$
        \sum_{i,j=1}^{n-1} \widetilde\lambda_{ij} = \sum_{i,j=1}^{n-1} \widehat\lambda_{ij} \qquad \text{and} \qquad \sum_{i,j=1}^{n-1} \widetilde\lambda_{ij}^{\, 2} = \widetilde{\sigma}^2.
    $$
\end{proposition}

Moreover, $\sum_{i,j} \widehat{\lambda}_{ij}^{\,2} > \widetilde{\sigma}^2 > (n-1)^2 \, \bar\lambda^2$
implies $\alpha \in (0,1)$ and hence $\widetilde\lambda_{ij} > 0$ for $i,j = 1,\ldots,n-1$;
otherwise no shrinkage is applied.

\medskip

\begin{remark}
The final step is to evaluate the upper tail probability of a Gaussian quadratic form $\sum_i \lambda_i Z_i^2$. This problem has been studied extensively and a variety of algorithms are available \citep{imhof1961,davies1980,ruben1962,liu2009,kuonen1999}. In our implementation, we use a hybrid approach, for which details are provided in Appendix B.1 in the Supplementary Material.
\end{remark}

\section{Simulations} \label{sec:simulations}

Throughout this simulation study, we will compare the spectral approach based on the test statistic $n \widehat{\V}_n^2$ with several competitors from the literature. \texttt{Gamma} denotes an approximation using a gamma distribution, choosing the scale and shape parameter of the distribution such that it matches estimates of the first and second moment of the test statistic (see, e.g. \cite{HH}); to maximize precision of this estimator, the unbiased estimators by Berschneider and Boettcher \cite{berschneider2018complex} are used. \texttt{BB2} and \texttt{BB3} denote two different approaches given in \cite{berschneider2018complex}. \texttt{BB2} is based on a gamma distribution and uses unbiased estimators of the first two moments of $n \widehat{\V}_n^2$; it is shown to be asymptotically valid and conservative for p-values smaller than 0.215. \texttt{BB3} is based on matching a Pearson type III distribution to estimates of the first three moments of $n \widehat{\V}_n^2$. \texttt{SPV} denotes the so-called \textit{chi-square test of distance correlation} by Shen, Panda and Vogelstein \cite{shen2022chi}. For completeness, we also evaluate the simple asymptotic approach provided in the original distance correlation by Sz\'{e}kely, Rizzo and Bakirov \cite{szekely2007measuring}, which will be denoted by \texttt{SRB}. As \texttt{BB2}, both \texttt{SPV} and \texttt{SRB} are known to be asymptotically valid and conservative for p-values smaller than 0.215.

\subsection{Computation time}

We first compare the computation time of different distance covariance tests. This additionally serves the purpose of demonstrating the efficiency of the adaptive eigenvalue algorithms proposed in Section~\ref{sec:bounds}. Since this algorithm is highly dependent on the true p-value (and hence the dependence of the underlying random variables) and the different accuracy parameters, we do not benchmark the adaptive scheme itself, but instead simply use an approximation based on the first $100$ eigenvalues of the two distance matrices, denoted by \texttt{Spectral (first 100)}. Additionally, we also show the computation time for the test based on the complete spectrum, which is denoted by \texttt{Spectral (Complete)}. Moreover, since the four algorithms \texttt{Gamma}, \texttt{BB2}, \texttt{SPV} and \texttt{SRB} essentially require the same steps and computation time, we omit the exact evaluation of the latter three approaches. 

For all computation time evaluations the random variables $X$ and $Y$ under consideration were independent univariate standard normally distributed. We considered two different versions of generalized distance covariance. The first setting investigates the classical distance covariance proposed in \cite{szekely2007measuring}, whereas the second uses the \textit{Gaussian} metric,
\begin{equation} \label{eq:gaussian}
    d(x_i,x_j) = 1 - \exp \big( -(x_i - x_j)^2/(2 h^2) \big),
\end{equation}
where the bandwidth $h$ is chosen using the median bandwidth heuristic. This test statistic corresponds to HSIC with a so-called Gaussian kernel \cite{sejdinovic2013equivalence, edelmann2022regression}, which is arguably the most popular kernel for HSIC in  machine learning. We considered the sample sizes $125, 250, 500, 1000, 2000, 4000, 8000, 16{,}000$ and $32{,}000$.
Exact details on the evaluation of the computation time are given in the Supplementary Material, Appendix B.2.

\begin{figure}[htbp]
  \centering
  \includegraphics[width=\textwidth]{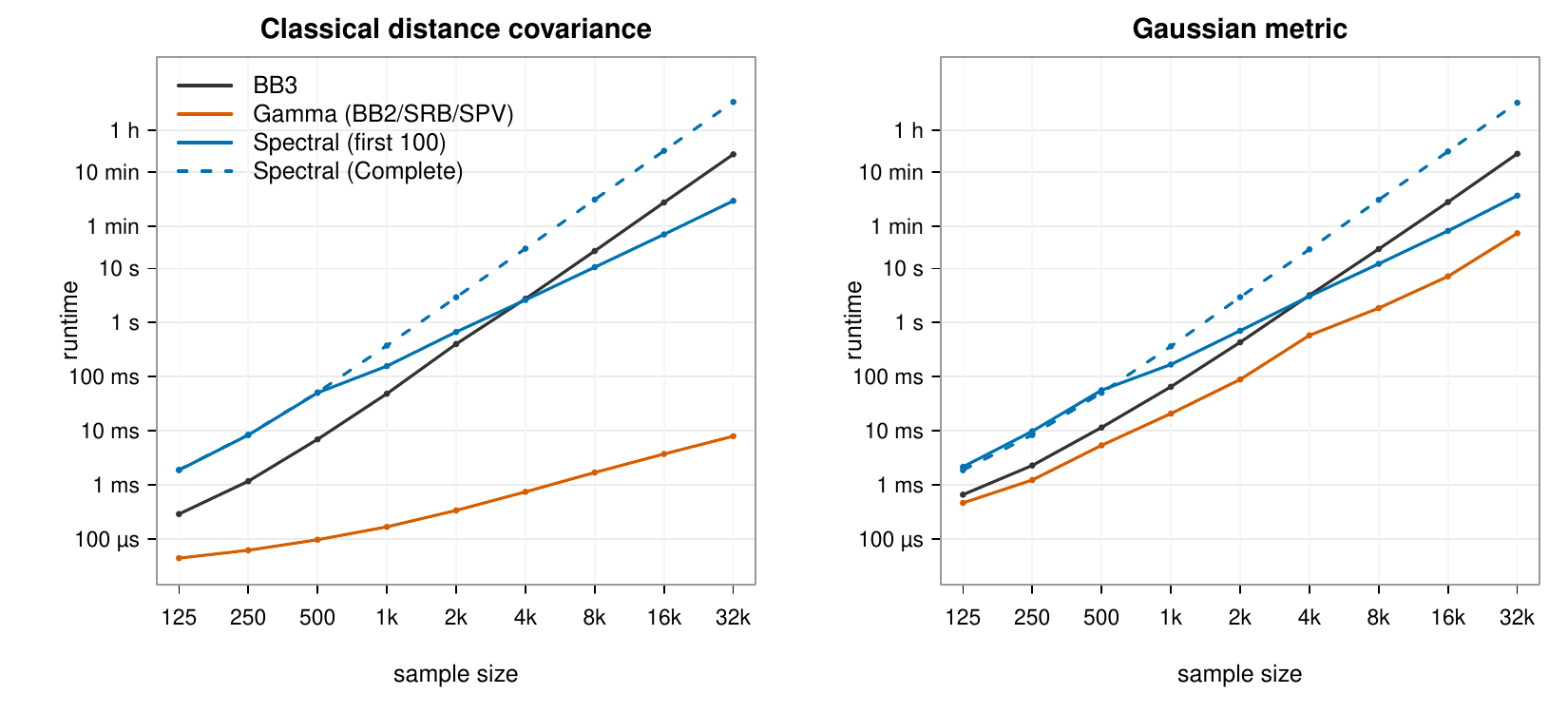}
  \caption{Runtime of different testing methods as a function of
    the sample size $n$. Left: classical distance covariance; right: generalized distance covariance with Gaussian metrics. Both axes are on a $\log_2$ scale.}
  \label{fig:runtime}
\end{figure}

The results are provided in Figure \ref{fig:runtime}, where we use logarithmic scales for both sample size and computation time. We note that the lines of the runtime for \texttt{BB3} and \texttt{Spectral (Complete)} are nearly parallel in both settings, which neatly illustrates that both procedures require $O(n^3)$ operations. While both procedures are reasonably fast for small and moderate sample sizes, they get computationally expensive for higher sample sizes; for $n = 32{,}000$, the runtime for \texttt{BB3} is $21.4$ minutes for the classical distance covariance and $22.0$ minutes for Gaussian metrics; for \texttt{Spectral (Complete)}  the corresponding runtimes are both greater than $3$ hours. By using a restricted eigenvalue decomposition, the runtime can be substantially reduced. While the runtime of \texttt{Spectral (first 100)} is equal to \texttt{Spectral (Complete)} for small sample sizes, since our implementation then defaults to calculate the complete eigendecomposition in any case, the runtime increases far slower with increasing sample sizes from then, due to the reduced complexity of $O(n^2)$; for $n = 32{,}000$, the runtime for \texttt{Spectral (first 100)} is just below $3$ minutes for the classical distance covariance and $3.7$ minutes for Gaussian metrics. While the runtime for two spectral methods and \texttt{BB3} is very similar for the two settings with slightly higher costs for the Gaussian metrics due to bandwidth selection, the runtimes are markedly different for \texttt{Gamma}. The reason is that our implementation then uses a $O(n \log n)$ algorithm \cite{huo2016fast} for the classical distance covariance of real-valued variables and estimates for the first two moments can be readily computed from terms arising from this algorithm (in contrast to \texttt{BB3}, which explicitly requires the third power of the distance matrices). The computational gain from this algorithm is enormous; even for a sample size of 1 million, the runtime is still below half a second. For the Gaussian metrics, such an algorithm is not available,  the runtime is $O(n^2)$ - and hence the corresponding line for the logarithmic runtime is - for larger sample sizes - nearly parallel to that of \texttt{Spectral (first 100)} and just by a factor of approximately $5$ to $10$ faster than this algorithm.

\subsection{Type I error}

In the second part of the simulation study, we compare the empirical type I error of the newly proposed spectral tests with their competitors. For these simulations, the spectral test is based on the complete eigenvalue decomposition. However, to investigate the precision gain achieved by the shrinkage approach  in Section \ref{sec:shrinkage}, we consider both the test with shrinkage and the naive approach using the ``plain spectrum'' without shrinkage. As in the first part of the simulation study, $X$ and $Y$  were independent and univariate standard normally distributed; both the classical distance covariance and generalized distance covariance using the Gaussian metric in \eqref{eq:gaussian} are investigated. The empirical type I error was calculated based on $10$ million simulations per scenario. For all methods, we evaluated the empirical type I error rate for the nominal levels $\alpha = 0.05, 0.005, 0.0005, 5 \times 10^{-5}, 5 \times 10^{-6}$. The smaller nominal levels are highly important when adjustments for multiple testing are necessary or extremely small significance levels are chosen (such as in genetics, cf. \cite{edelmann2025generalized}). We considered the sample sizes $n = 10, 20, 50, 100, 200, 500$.
The results for the classical distance covariance are given in Table \ref{tab:classical}, the results for the generalized distance covariance with Gaussian metrics are given in Table \ref{tab:gaussian}. For easier readability, the results for the empirical type I error are given as factors of the corresponding nominal level $\alpha$.

Except for \texttt{BB2} and \texttt{SRB}, that are markedly conservative, all methods give  satisfactory results for $\alpha = 0.05$. However, while the two spectral tests converge to the nominal level (as theoretically shown in Section \ref{sec:main}), the empirical type I error of the other tests appears to converge to ``some value close to the nominal level''. For smaller nominal levels, \texttt{Gamma} gets increasingly anti-conservative (with type I error inflation up to a factor of $72.06$), whereas \texttt{SPV} gets increasingly conservative, while - at least for larger sample sizes - the empirical type I error of \texttt{BB3} and the spectral tests are not very far from $\alpha$. 

In general \texttt{BB3} typically shows the best performance for small sample sizes ($n \leq 50$), whereas the spectral tests show the best performance for larger sample sizes ($n > 50$). \texttt{Spectral (shrinkage)} is better than \texttt{Spectral (naive)} in almost every scenario, giving a numerical justification of the shrinkage approach developed in Section \ref{sec:shrinkage}.

While it is clearly outperformed by \texttt{BB3} and the spectral approaches, we also note that the \texttt{SPV} approach shows a surprisingly good performance for standard distance covariance. Considering that it can be calculated in merely $O(n \log n)$ operations and is asymptotically conservative, it may be useful in certain applications.
On the other hand - although, we don't have theoretical evidence - the \texttt{Gamma} approach appears to be markedly anticonservative and may - in settings with many tested hypotheses - serve as a filter to select hypotheses to evaluate with a more precise spectral approach.

\begin{table}[ht]
\centering
\caption{Rejection rates as factors of $\alpha$ under $H_0$ for classical distance covariance. For each sample size, the entry closest to the nominal level is given in bold; entries equal to $0.00$, corresponding to no rejection in any of the simulation runs, are excluded from this comparison.}
\label{tab:classical}
\begin{tabular}{lrrrrrr}
\toprule
Method & $n=10$ & $n=20$ & $n=50$ & $n=100$ & $n=200$ & $n=500$ \\
\midrule
\multicolumn{7}{l}{\textbf{$\alpha = \text{0.05}$}} \\
Gamma & 1.20 & 1.12 & 1.09 & 1.08 & 1.07 & 1.08 \\
Spectral (shrinkage) & 1.11 & 1.04 & 1.02 & \textbf{1.01} & \textbf{1.00} & 1.01 \\
Spectral (naive) & 0.79 & 0.90 & 0.96 & 0.98 & 0.99 & \textbf{1.00} \\
BB3 & \textbf{1.10} & \textbf{0.99} & 0.95 & 0.94 & 0.93 & 0.94 \\
BB2 & 0.15 & 0.20 & 0.22 & 0.22 & 0.23 & 0.23 \\
SRB& 0.01 & 0.01 & 0.02 & 0.02 & 0.02 & 0.02 \\
SPV & 1.44 & 1.12 & \textbf{0.99} & 0.96 & 0.94 & 0.93 \\
\midrule
\multicolumn{7}{l}{\textbf{$\alpha = \text{0.005}$}} \\
Gamma & 2.05 & 2.37 & 2.47 & 2.49 & 2.49 & 2.52 \\
Spectral (shrinkage) & 0.42 & 0.78 & \textbf{0.94} & \textbf{0.97} & \textbf{0.97} & \textbf{1.00} \\
Spectral (naive) & 0.03 & 0.45 & 0.78 & 0.89 & 0.93 & 0.98 \\
BB3 & \textbf{1.08} & \textbf{1.04} & \textbf{1.06} & 1.06 & 1.05 & 1.07 \\
BB2 & 0.00 & 0.03 & 0.07 & 0.08 & 0.09 & 0.09 \\
SRB& 0.00 & 0.00 & 0.00 & 0.00 & 0.00 & 0.00 \\
SPV & 0.66 & 0.71 & 0.69 & 0.68 & 0.67 & 0.68 \\
\midrule
\multicolumn{7}{l}{\textbf{$\alpha = \text{0.0005}$}} \\
Gamma & 1.90 & 5.26 & 6.56 & 6.90 & 6.97 & 7.19 \\
Spectral (shrinkage) & 0.00 & 0.29 & 0.72 & \textbf{0.86} & \textbf{0.91} & \textbf{0.97} \\
Spectral (naive) & 0.00 & 0.08 & 0.52 & 0.75 & 0.85 & 0.94 \\
BB3 & \textbf{0.79} & \textbf{0.81} & \textbf{1.14} & 1.25 & 1.28 & 1.32 \\
BB2 & 0.00 & 0.00 & 0.02 & 0.03 & 0.04 & 0.05 \\
SRB& 0.00 & 0.00 & 0.00 & 0.00 & 0.00 & 0.00 \\
SPV & 0.00 & 0.19 & 0.40 & 0.45 & 0.47 & 0.47 \\
\midrule
\multicolumn{7}{l}{\textbf{$\alpha = \text{5e-05}$}} \\
Gamma & 0.24 & 10.70 & 18.41 & 20.82 & 21.40 & 22.39 \\
Spectral (shrinkage) & 0.00 & 0.02 & 0.48 & \textbf{0.66} & \textbf{0.78} & \textbf{0.89} \\
Spectral (naive) & 0.00 & 0.00 & 0.28 & 0.54 & 0.70 & 0.85 \\
BB3 & \textbf{1.23} & \textbf{0.36} & \textbf{1.21} & 1.38 & 1.58 & 1.69 \\
BB2 & 0.00 & 0.00 & 0.00 & 0.01 & 0.01 & 0.01 \\
SRB& 0.00 & 0.00 & 0.00 & 0.00 & 0.00 & 0.00 \\
SPV & 0.00 & 0.01 & 0.17 & 0.26 & 0.31 & 0.34 \\
\midrule
\multicolumn{7}{l}{\textbf{$\alpha = \text{5e-06}$}} \\
Gamma & 0.00 & 18.70 & 52.92 & 64.66 & 68.38 & 72.06 \\
Spectral (shrinkage) & 0.00 & 0.00 & 0.20 & \textbf{0.60} & \textbf{0.58} & \textbf{0.62} \\
Spectral (naive) & 0.00 & 0.00 & 0.14 & 0.42 & 0.48 & 0.60 \\
BB3 & \textbf{2.10} & \textbf{0.08} & \textbf{0.94} & 1.56 & 1.96 & 2.02 \\
BB2 & 0.00 & 0.00 & 0.00 & 0.00 & 0.00 & 0.00 \\
SRB& 0.00 & 0.00 & 0.00 & 0.00 & 0.00 & 0.00 \\
SPV & 0.00 & 0.00 & 0.04 & 0.08 & 0.14 & 0.10 \\
\bottomrule
\end{tabular}
\end{table}

\begin{table}[ht]
\centering
\caption{Rejection rates as factors of $\alpha$ under $H_0$ for generalized distance covariance with Gaussian distance and median bandwidth heuristic. Bold entries as in Table~\ref{tab:classical}.}
\label{tab:gaussian}
\begin{tabular}{lrrrrrr}
\toprule
Method & $n=10$ & $n=20$ & $n=50$ & $n=100$ & $n=200$ & $n=500$ \\
\midrule
\multicolumn{7}{l}{\textbf{$\alpha = \text{0.05}$}} \\
Gamma & 1.12 & 1.08 & 1.07 & 1.06 & 1.06 & 1.06 \\
Spectral (shrinkage) & 1.11 & 1.05 & \textbf{1.02} & \textbf{1.01} & \textbf{1.00} & \textbf{1.00} \\
Spectral (naive) & 0.79 & 0.90 & 0.96 & 0.98 & 0.99 & 0.99 \\
BB3 & \textbf{1.07} & \textbf{1.00} & \textbf{0.98} & 0.97 & 0.97 & 0.97 \\
BB2 & 0.10 & 0.12 & 0.13 & 0.13 & 0.13 & 0.14 \\
SRB& 0.01 & 0.00 & 0.00 & 0.00 & 0.00 & 0.00 \\
SPV & 1.39 & 1.08 & 0.95 & 0.91 & 0.89 & 0.88 \\
\midrule
\multicolumn{7}{l}{\textbf{$\alpha = \text{0.005}$}} \\
Gamma & 1.25 & 1.53 & 1.63 & 1.65 & 1.67 & 1.67 \\
Spectral (shrinkage) & 0.41 & 0.79 & 0.94 & \textbf{0.97} & \textbf{0.99} & \textbf{1.00} \\
Spectral (naive) & 0.03 & 0.45 & 0.78 & 0.89 & 0.95 & 0.98 \\
BB3 & \textbf{0.91} & \textbf{0.97} & \textbf{1.02} & 1.04 & 1.04 & 1.05 \\
BB2 & 0.00 & 0.01 & 0.01 & 0.02 & 0.02 & 0.02 \\
SRB& 0.00 & 0.00 & 0.00 & 0.00 & 0.00 & 0.00 \\
SPV & 0.26 & 0.31 & 0.29 & 0.28 & 0.28 & 0.27 \\
\midrule
\multicolumn{7}{l}{\textbf{$\alpha = \text{0.0005}$}} \\
Gamma & \textbf{0.57} & 2.14 & 2.80 & 3.00 & 3.09 & 3.12 \\
Spectral (shrinkage) & 0.00 & 0.32 & 0.73 & \textbf{0.88} & \textbf{0.95} & \textbf{0.97} \\
Spectral (naive) & 0.00 & 0.09 & 0.52 & 0.76 & 0.89 & 0.94 \\
BB3 & 0.54 & \textbf{0.77} & \textbf{1.09} & 1.21 & 1.24 & 1.26 \\
BB2 & 0.00 & 0.00 & 0.00 & 0.00 & 0.00 & 0.00 \\
SRB& 0.00 & 0.00 & 0.00 & 0.00 & 0.00 & 0.00 \\
SPV & 0.00 & 0.03 & 0.06 & 0.08 & 0.07 & 0.06 \\
\midrule
\multicolumn{7}{l}{\textbf{$\alpha = \text{5e-05}$}} \\
Gamma & 0.06 & 2.71 & 5.12 & 6.00 & 6.22 & 6.29 \\
Spectral (shrinkage) & 0.00 & 0.02 & 0.49 & \textbf{0.71} & \textbf{0.88} & \textbf{0.92} \\
Spectral (naive) & 0.00 & 0.00 & 0.28 & 0.57 & 0.78 & 0.88 \\
BB3 & \textbf{0.47} & \textbf{0.38} & \textbf{1.18} & 1.51 & 1.54 & 1.60 \\
BB2 & 0.00 & 0.00 & 0.00 & 0.00 & 0.00 & 0.00 \\
SRB& 0.00 & 0.00 & 0.00 & 0.00 & 0.00 & 0.00 \\
SPV & 0.00 & 0.00 & 0.01 & 0.02 & 0.02 & 0.02 \\
\midrule
\multicolumn{7}{l}{\textbf{$\alpha = \text{5e-06}$}} \\
Gamma & 0.00 & 2.44 & 9.70 & 12.80 & 13.30 & 13.96 \\
Spectral (shrinkage) & 0.00 & 0.00 & 0.32 & \textbf{0.66} & \textbf{0.84} & \textbf{0.84} \\
Spectral (naive) & 0.00 & 0.00 & 0.18 & 0.40 & 0.66 & 0.78 \\
BB3 & \textbf{0.54} & \textbf{0.20} & \textbf{1.20} & 2.08 & 2.32 & 2.22 \\
BB2 & 0.00 & 0.00 & 0.00 & 0.00 & 0.00 & 0.00 \\
SRB& 0.00 & 0.00 & 0.00 & 0.00 & 0.00 & 0.00 \\
SPV & 0.00 & 0.00 & 0.00 & 0.00 & 0.02 & 0.00 \\
\bottomrule
\end{tabular}
\end{table}

\section{Discussion} \label{sec:discussion}

In this article we have studied the direct approximation of the null distribution of a generalized distance covariance through the spectra of the doubly centered distance matrices of the respective samples. To the best of our knowledge, we have provided the first
rigorous justification that the empirical spectra of $A_n$ and $B_n$ yield a consistent approximation of the limiting distribution and hence an asymptotically valid test. 
Moreover, we have derived novel algorithms for these tests that improve both the precision and runtime compared to naive implementations. Using a simulation study we provided numerical evidence for our findings and investigated the performance of our algorithms. While for moderate sample sizes $n \ge 100$, the precision of our spectral tests outperforms all competitors, we also note that, for very small samples ($n \le 50$), the performance is rather weak. For such small sample sizes, the spectral tests should hence not be used - the obvious choice in this case is a permutation procedure, which is not only exact, but also very fast in the small sample setting. If for some reason, a Monte-Carlo procedure is not desired, the three-moment based approximation \texttt{BB3} should be preferred.

Several limitations point to possible future extensions of this work. Except for a few special cases, all algorithms get prohibitively expensive in terms of runtime and memory requirements for very large sample sizes $n$ due to the problem of dealing with $n \times n$ matrices. There are several potential approaches to mitigate this issue. By rounding or binning the observations in $\bX$ and $\bY$, one may reduce the number of distinct values to $m \ll n$; the algorithm then merely requires two $m \times m$ distance matrices and one $m \times m$ contingency table. Another approach may be to approximate the kernels implied by the distances \cite{edelmann2022regression} by a finite number of $k$ features, reducing the computational complexity to $k^2 n$. The weak performance of the spectral algorithms for very small sample sizes may be improved by finding ways to incorporate estimates of the third moment into the spectral approaches - one may even hope that this leads to a uniform improvement over both methods and hence also better performance for larger sample sizes.

More broadly, because the generalized distance covariance studied here coincides
with HSIC for the corresponding kernels
\cite{sejdinovic2013equivalence,edelmann2022regression}, the results transfer
directly to kernel-based independence testing. Beyond this, one may expect that
the theoretical derivation of our approximation results extends in a
straightforward manner to many other degenerate $U$- and $V$-statistics; similarly the adaptive spectral scheme and the shrinkage correction developed here should be of use well beyond
the setting of distance covariance.

\bibliographystyle{plain} % Style BST file (imsart-number.bst or imsart-nameyear.bst)
\bibliography{approx}

\end{document}